\documentclass[%
 reprint,
 amsmath,amssymb,
 aps,
]{revtex4-2}

\usepackage{siunitx} 
\usepackage{graphicx}
\usepackage{dcolumn}
\usepackage{bm}
\usepackage{gensymb}
\usepackage{verbatim} 

\begin{document}

\preprint{APS/123-QED}

\title{Entanglement in living systems}

\author{Thomas C. Day}

\author{S. Alireza Zamani-Dahaj}

\author{G. Ozan Bozdag}

\author{Anthony J. Burnetti}

\author{Emma P. Bingham}

\author{Peter L. Conlin}

\author{William C. Ratcliff}

\author{Peter J. Yunker}

\date{\today}

\begin{abstract}

Many organisms exhibit branching morphologies that twist around each other and become entangled. Entanglement occurs when different objects interlock with each other, creating complex and often irreversible configurations. This physical phenomenon is well-studied in non-living materials, such as granular matter, polymers, and wires, where it has been shown that entanglement is highly sensitive to the geometry of the component parts. However, entanglement is not yet well understood in living systems, despite its presence in many organisms. In fact, recent work has shown that entanglement can evolve rapidly, and play a crucial role in the evolution of tough, macroscopic multicellular groups. Here, through a combination of experiments, simulations, and numerical analyses, we show that growth generically facilitates entanglement for a broad range of geometries. We find that experimentally grown entangled branches can be difficult or even impossible to disassemble through translation and rotation of rigid components, suggesting that there are many configurations of branches that growth can access that agitation cannot. We use simulations to show that branching trees readily grow into entangled configurations. In contrast to non-growing entangled materials, these trees entangle for a broad range of branch geometries. We thus propose that entanglement via growth is largely insensitive to the geometry of branched-trees, but instead will depend sensitively on time scales, ultimately achieving an entangled state once sufficient growth has occurred. We test this hypothesis in experiments with snowflake yeast, a model system of undifferentiated, branched multicellularity, showing that lengthening the time of growth leads to entanglement, and that entanglement via growth can occur for a wide range of geometries. Taken together, our work demonstrates that entanglement is more readily achieved in living systems than in their non-living counterparts, providing a widely-accessible and powerful mechanism for the evolution of novel biological material properties.

\end{abstract}

\maketitle

\begin{figure*}
    \includegraphics{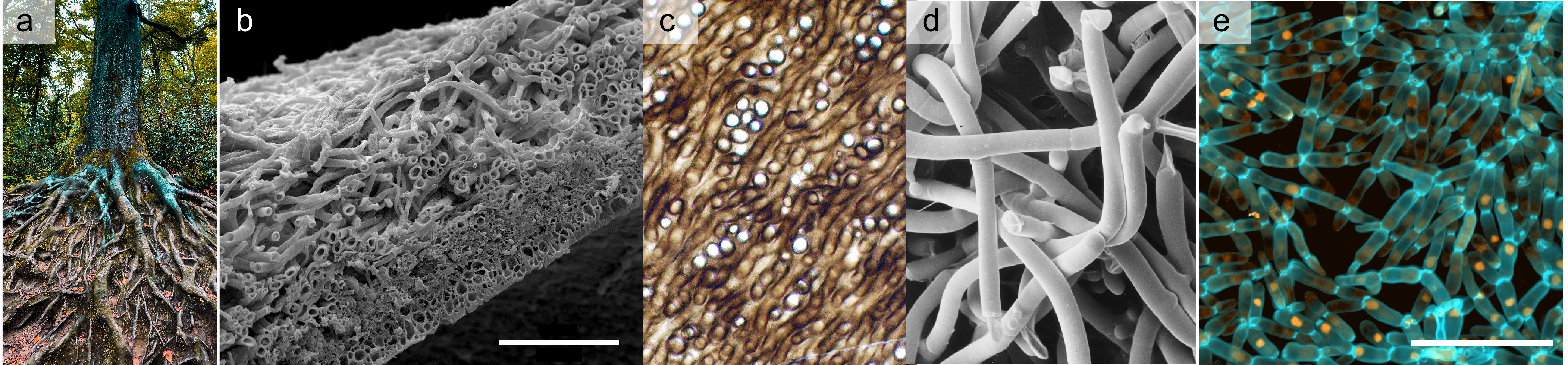}
    \caption{\textbf{Several examples of entangled, growing materials}. \textbf{a} Tree roots winding and twisting around each other. Photo used with permission from Omar Ram via Unsplash. \textbf{b} \textit{P. membranacea}, a type of lichen, in cross section. Scale bar is \SI{100}{\micro\meter}. Image source: $https://cstremblog.blogspot.com/2018/03/peltigera-membranacea-cryofracture.html$. Image used with permission from Chistopher Tomellion. \textbf{c} The fossilized (probable) fungus \textit{Prototaxites}, which formed structures 8 meters tall 400 million years ago. Strands are about \SI{50}{\micro\meter} in diameter. Image Source: $https://commons.wikimedia.org/wiki/File:Prototaxm10.JPG$. This image is in the public domain. \textbf{d} Scanning electron micrograph of hyphae of \textit{Pleurotus}. The hyphae have a diameter of about \SI{3}{\micro\meter}. Image Source: $http://www.davidmoore.org.uk/Sec01_03.htm$. Image used with permission from Dr. Carmen Sanchez. \textbf{e} Confocal microscope image of snowflake yeast, scale bar \SI{50}{\micro\meter}.}
    \label{FIG_01}
\end{figure*}

Many organisms grow with filamentous, branching morphologies, including plants, mycelial networks, cyanobacterial mats, and more. These branched tree-like organisms often wind around themselves or others, thus becoming visually tangled (Figure \ref{FIG_01}). This physical phenomenon, called ``entanglement'', is well-studied in non-living materials \cite{Raymer2007, Gravish2012, Brown2012, Tang2014, Meng2018, Edwards2020, Norioka2021, Kim2021, Marenduzzo2010, Meluzzi2010}, and has recently become a topic of interest in active systems \cite{Hu2016, Becker2022, Patil2023}. Entanglement provides these systems unique and potentially useful material properties. For instance, materials composed of entangled components are generally both strong and tough \cite{Edwards2020, Kim2021} and exhibit strain stiffening \cite{Brown2012}. These studies also make it clear that entanglement requires precise engineering of the structure and geometry of the entangling constituents \cite{Gravish2012, Tang2014, Edwards2020, Norioka2021, Kim2021}. However, the growth of an organism is qualitatively distinct from the assembly of non-living materials. Living systems experience birth and death events, providing sink and source terms to their continuity equation, and are also evolved rather than designed. Therefore, the rules for generating non-growing entangled materials do not necessarily apply to entanglement via growth, leaving it unclear what determines whether growing systems do or do not entangle.

It was recently discovered that entanglement rapidly evolves, \textit{de novo}, in multicellular yeast clusters \cite{Bozdag2023}. These clusters, known as ``snowflake yeast'', initially grow as branched trees. They were subjected to selection for large size every day for 600 days; over this time, snowflake yeast evolved a new morphology in which disconnected branches are physically entangled, enabling clusters to grow larger than \SI{1}{\milli\meter} in size. The speed and ease with which snowflake yeast evolved entanglement, combined with the presence of many entangled organisms in nature (Figure \ref{FIG_01}), suggests that either all of these organisms are coincidentally positioned near a specific structural and geometric entanglement sweet spot, or that there is a broader physical principle that enables entanglement via growth for a wide range of branched trees.

Here, we use a combination of experiments and a variety of numerical modeling methods to show that growth readily establishes entanglement in branched trees for nearly any geometry, unlike entanglement of non-growing elements. We find that these entangled configurations are difficult or even impossible to access through translation and rotation alone, suggesting that entanglement from growth is fundamentally distinct from entanglement absent growth (i.e., non-living materials). First, we use numerical manipulations of experimental data to interrogate what kinds of entangled branches can or cannot be disassembled. Then, we develop a simple simulation to investigate how entangled configurations of branches arise, and how entanglement probability is affected by geometric properties of the branches. Surprisingly, we find that entanglement via growth is generically easy to achieve, almost regardless of branch geometry. This led us to develop a simple model, without specifying a growth morphology, to explore the onset of entanglement via growth. We find that entanglement can be a slow process, suggesting that for growing branched trees, entanglement depends primarily on time scales --- if growth does or does not stop before entanglement is complete --- rather than geometry. We test this idea in experiments by growing branching microbes, manipulating explicitly the time for which they sit next to one another, and separately their branching geometry, confirming that time scales control entanglement via growth. 

\section*{Entangled, growing branches}
\begin{figure*}
    \centering
    \includegraphics[width=\linewidth]{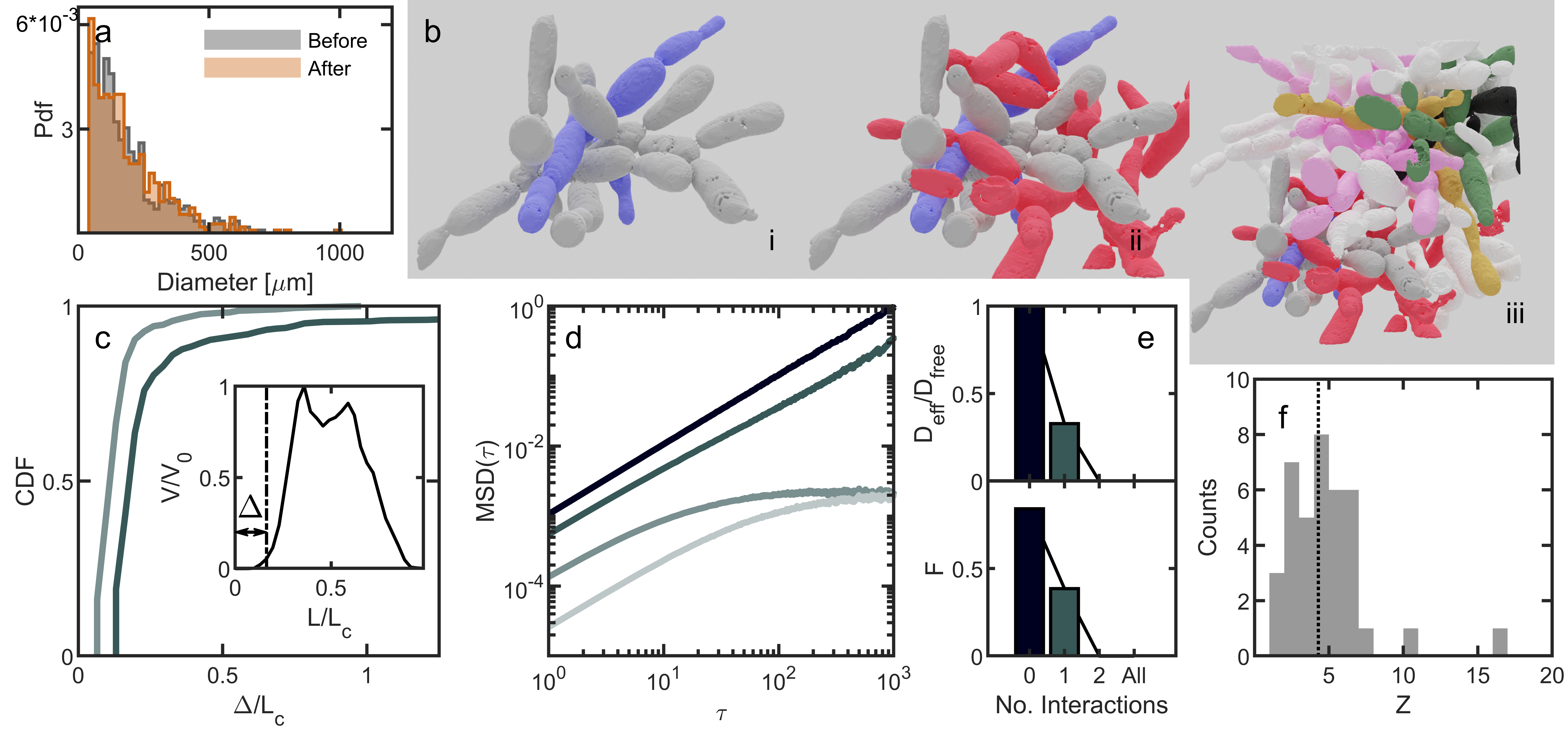}
    \caption{\textbf{Growing branches access configurations inaccessible or difficult to access through agitation alone.} \textbf{a} Histogram of yeast group sizes before and after strong agitation via vortexing. \textbf{b} Several examples of entangled branches. (i) Two pieces penetrate each other's empty space, (ii) a third piece also entangles with the previous two, (iii) a view of all pieces identified in the sample data cube. \textbf{c} Cumulative distribution function for the distance dragged until the point of first contact, in units of distance scaled by average cell length. Dark line, two piece interaction from a.i, lighter line is the 3 piece interaction from a.ii. Inset: an example of one drag run, showing net overlap scaled by the maximum overlap and distance pulled scaled by cell length. The distance to defined first contact, $\Delta$, is illustrated. \textbf{d} Mean-squared displacement vs. lag time for four different agitation interaction scenarios. The target piece is always the gray piece from \textbf{b}. From dark to light, the lines represent free diffusion, one interaction (b.i), two interactions (b.ii), and all interactions (b.iii). \textbf{e}, Top is the effective diffusion constant for all lines from \textbf{d}, scaled by the free diffusion constant. Bottom, the fraction of independent simulations that translated at least 1 cell length. \textbf{f} The coordination number for all 38 pieces from b.iii.}
    \label{FIG_02}
\end{figure*}

We begin by investigating an experimental system that is known to grow into entangled configurations, a multicellular Baker's yeast called snowflake yeast \cite{Bozdag2023}. Snowflake yeast form structures that resemble branching trees via continued rounds of cell division. New cells bud from their mother cell and remain attached through a rigid chitinous bond; if the bond breaks, it is not reformable. Cells do not adhere via sticky interactions such as surface flocculation proteins or extracellular matrix. Therefore, cells are connected one to another in a tree-like pattern, such that breaking any chitinous bond breaks the group into two pieces \cite{Ratcliff2012, Ratcliff2015, Jacobeen2018a}. We used snowflake yeast strains taken from an ongoing long-term evolution experiment \cite{Bozdag2023}. We have previously found that branches of yeast cells can interact sterically with one another, intercalating and entangling within a single yeast tree \cite{Bozdag2023}, with entanglement arising \textit{de novo} in less than 600 days of experimental evolution. Given the precision necessary to create non-living entangled materials, it is surprising that this new morphology evolved so readily in all five independently evolving populations. This observation suggests that perhaps entanglement via growth is fundamentally different than entanglement of non-living materials.

We first tested if agitation affects the integrity of entangled branches. Unlike previous experiments with entangled granular materials, in which mechanical agitation leads to collapse of a rigid column \cite{Gravish2012}, or in observations of active tangled matter, which can reversibly tangle and untangle quickly \cite{Patil2023}, the snowflake yeast branches appear difficult, if not impossible, to disassemble. When vortex mixed at medium strength, snowflake yeast groups maintain their size distribution, suggesting that mechanical agitation alone (weak enough to not break intercellular bonds) cannot disassemble the tangled aggregate (Figure \ref{FIG_02}a). Crucially, unlike knot and tangle theories, there is no rigorous theory of rigid-body entanglement with loose ends, so there is no theoretical test that can assess if two or more rigid, free-ended objects can or cannot be dis-entangled. We therefore turned to empirical tests of snowflake yeast at the micro-scale to confirm if growth accesses configurations that cannot be disassembled.

Previous work used scanning electron microscopy to image 3D volumes of a single snowflake yeast group \cite{Bozdag2023}. Here, we segmented each disconnected branch of cells in those 3D image stacks, and generated 3D surface data by approximating their surfaces through alpha shapes (Figure \ref{FIG_02}b). We then computationally simulated artificial translations and rotations of various branches of cells and tracked their collisions (see Methods). Using this method, we investigated if the experimentally observed configurations that snowflake yeast grew into could be disassembled via mechanical agitation.


We quantified the degree of confinement by performing simulations in which we translated one yeast branch with respect to others. We identified two separate branches (Figure \ref{FIG_02}b.i) that were entangled, where entanglement was defined as occurring when one disconnected branch penetrated the convex hull of another, a definition broadly used when studying entanglement \cite{Brown2012, Gravish2012,Bozdag2023}. We constructed alpha shapes of both pieces. Then, we translated the two alpha shapes with respect to each other by identifying one target piece (the gray piece in Figure \ref{FIG_02}b) and one stationary piece and moving the target piece a distance of $1.3$ cell lengths in discrete steps of size $0.03$ cell lengths. We allowed the alpha shapes to overlap, and at each step measured the overlapping volume between the two alpha shapes. We repeated this ``drag experiment'' 1000 times in different directions, each direction defined by a unit direction vector, each vector evenly dispersed around the unit sphere. Contact is defined to be the point at which the overlapping volume exceeded one cubic micron (see Methods). Of the 1000 sampled directions, 35 did not make contact exceeding this threshold, indicating that the two pieces are not prohibitively entangled. The median first contact distance was $0.20$ cell lengths (Figure \ref{FIG_02}c). We next added a third branch of cells from the same snowflake yeast cluster (Figure \ref{FIG_02}b.ii) and repeated the drag simulation. With three branches, contact occurs in every translation direction, and the median first contact distance was $0.13$ cell lengths. Thus, in this example, the entangled piece can sometimes escape one neighbor, but it cannot escape two neighbors.

While these drag simulations suggest that these three branches are highly entangled, it is possible that the target branch can escape with a simple series of translations and rotations---maneuvers that are readily accessible to non-living materials. To test this idea, we randomly translated and rotated the branches to determine if they can undo snowflake yeast entanglement from growth. In our algorithm, one branch (the target) experienced movements that combine a random translation (a step of $\SI{0.4}{\micro\meter}$, or $\sim 0.03$ cell lengths, in a random direction) and a random rotation (a rotation of 2 degrees around a randomly selected axis). Collisions were identified by tracking the overlapping volume of the target branch with other branches. Random movements were accepted if the branches did not collide, and rejected if they did collide, in which case the target piece remained at its last non-overlapping position and orientation. From our drag experiments, we hypothesized that our target branch could be disassembled if it only interacted with one other branch, but may be confined when interacting with two or more others.

Following this procedure, we first agitated the target branch (gray, Figure \ref{FIG_02}b.i) in free space, tracking the position of its center of mass and calculating its mean squared displacement (MSD, $\langle( x(t+\tau) - x(t) )^2\rangle$) over 100 simulations, each running for 3000 time steps. The unconfined branch moved diffusively with diffusion constant $D_{0}=9.77*10^{-4} \pm 1*10^{-6}$ cell lengths squared per simulated time step. Next, we simulated a pair of interacting branches (Figure \ref{FIG_02}b.i) and ran $100$ replicate simulations. We used the same target branch (gray) as for the freely diffusing case. We found that the target branch still moved diffusively, which is consistent with our previous observation that the two-piece interaction is escapable. However, the effective diffusion constant was lower ($0.33D_0 = 3.21*10^{-4}\pm 1*10^{-6}$ cell lengths squared per unit time) due to the many collisions between the two pieces. Upon adding a third disconnected branch (Figure \ref{FIG_02}b.ii), we found that the MSD of the target piece ceases to grow linearly, indicating that it is caged by its neighbors. Upon adding all remaining pieces (Figure \ref{FIG_02}b.iii), motion was even more limited. To quantify this caging effect, we measured an effective diffusion constant for all four scenarios with the gray target piece, scaled by the free-space diffusion constant, and found that $D_{eff}$ approaches zero for three and four branch simulations (Figure \ref{FIG_02}e, top). Further supporting the caging observations, we found that the fraction of agitation simulations for which the target piece moved at least one cell length scales with the effective diffusion constant (Figure \ref{FIG_02}e, bottom, Pearson correlation coefficient $r=0.86$).

To test if other branches in the cluster behave similarly, we repeated this agitation experiment with an entirely different set of branches (Figure \ref{FIG_02}b.iii). In this example, we agitated the yellow branch with zero interactions, one interacting branch (pink), two interacting branches (pink and green), and three interacting branches (pink, green and black). In supplemental Figure \ref{SF_01} we report the same characteristic flattening of the mean squared displacement upon adding the second interaction, and include measurements of the effective diffusion constants. Last, as a demonstration for just how dramatic this caging effect can be, we agitated one branch that was entangled with 16 others (pink, Figure \ref{FIG_02}b.iii). After 50 replicate simulations, each with 3000 timesteps, the agitation algorithm was never successful in completing even a single accepted move (i.e. one that resulted in zero collisions).

The above results suggest that branches entangled with two or more other branches grew into highly confined configurations that would be very difficult, if not impossible, to reach through translations and rotations. We thus next sought to determine how many branches are entangled with two or more other branches within macroscopic clusters. To do so, we identified 38 discrete branches and computed the convex hull of each one. Then, for each component, we determined how many other convex hulls it penetrated, i.e., its coordination number, $z$ (Figure \ref{FIG_02}f). We found that all branches penetrate the convex hull of at least one other branch, and $92\%$ of the branches penetrate the convex hulls of two or more other pieces. The average coordination number was $\langle z\rangle=4.2\pm2.7$. Therefore, snowflake yeast branches appear to be highly confined.

These analyses of entangled branches suggest that entanglement via growth can achieve configurations that are difficult, if not impossible, to disassemble via translation and rotation alone. In these configurations, the only way to disassemble two or more entangled branches appears to be to destroy or deform the material, for example through external forces that rupture cell-cell bonds or via branch death. However, it is unclear if snowflake yeast coincidentally possessed a growth morphology with a geometry conducive to such highly confined, entangled branches, or if entanglement via growth is readily able to access such configurations. To answer this question, we sought to explore entanglement through growth via a model system that grows with a branched morphology and a tunable geometry.

\section*{Entanglement from growth via rigid-body simulations}
\begin{figure*}
    \includegraphics[width=\linewidth]{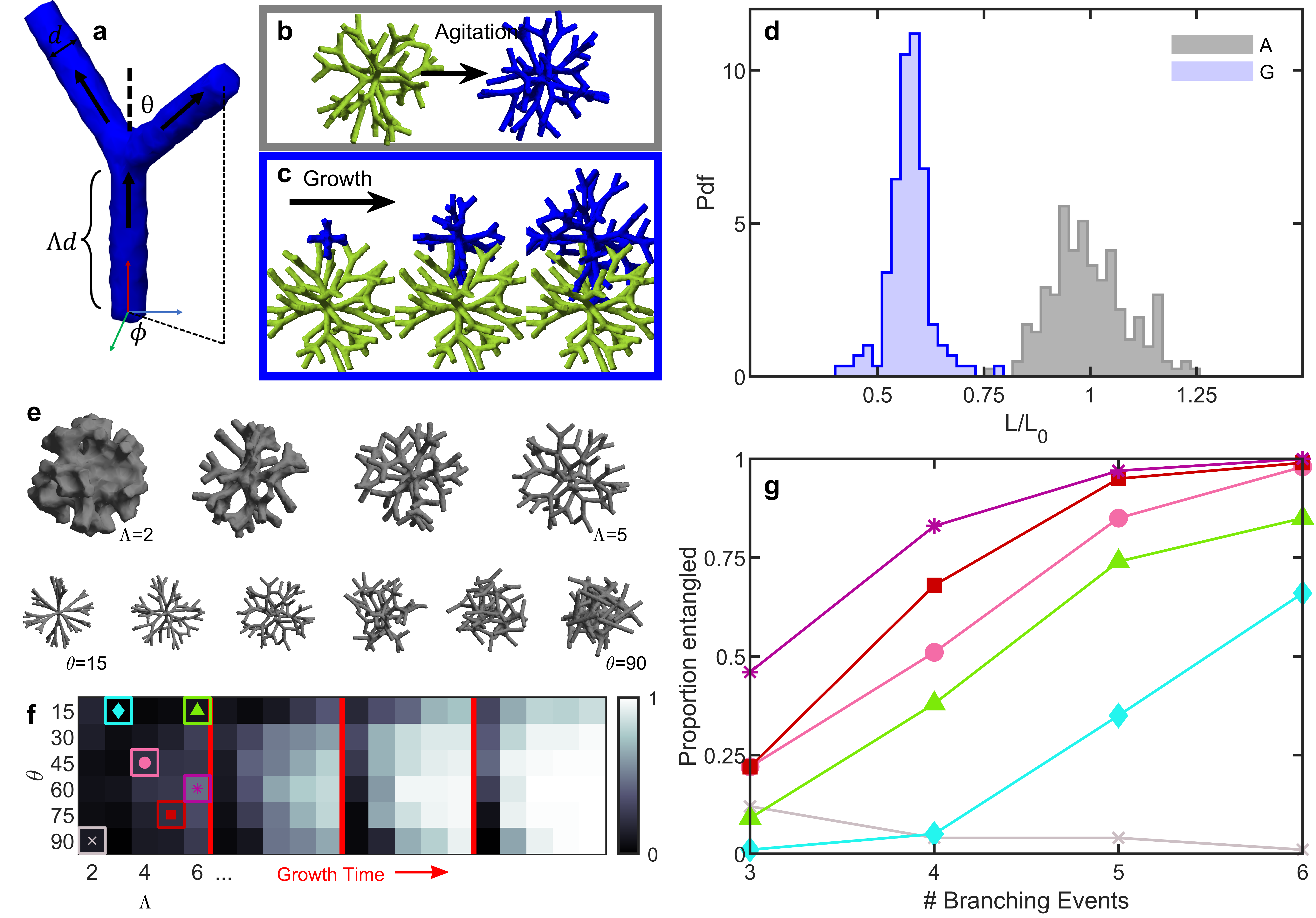}
    \caption{\textbf{Simulations of growing branches easily entangle.} \textbf{a} Illustration of a growing hyphal branch with one tip splitting into two tips, along with relevant geometric parameters. \textbf{b} Two separately-generated trees are pushed together and mechanically agitated. \textbf{c} One tree is first generated, and then a second is grown nearby (3 stages of growth are shown - early, middle and late times). \textbf{d} Histograms of distances between pairs of trees centers of mass. `A' stands for trees grown separately and agitated, `G' stands for trees grown nearby. Lengths are scaled by the mean distance achieved from agitation alone. \textbf{e} Examples of individual grown trees with varying geometric parameters. Top row: varying $\Lambda$ from $2-5$, bottom row: varying $\theta$ from $15-90\degree$. \textbf{f} Phase maps measuring the proportion of pairs of trees that are measured as entangled. Branching geometry is varied for 4 different growth times. From left to right: $3$ branching events (short times), $4$ branching events (intermediate), $5$, and $6$. \textbf{g} Tracking trajectories of entanglement probability for the specific branching geometries highlighted in \textbf{f}.}
    \label{FIG_03}
\end{figure*}

To test if entanglement is, in general, readily achieved via growth, we simulated growing, branching trees in three dimensions with a variety of geometries. We varied the geometry of these growing trees and determined which geometries do and do not allow entanglement to occur. Our simulations start with $6$ ``primary'' tips, centered at the origin, with each tip pointed along one of the cardinal axes. Tips have a fixed diameter, $d$, and grow by continually lengthening at a constant rate. After lengthening by a distance $\Lambda d$, each tip branches into two tips. The two new tips branch symmetrically from the growth axis prior to splitting, with branching angle $\theta$ and random azimuthal orientation $\phi$ (Figure \ref{FIG_03}a). If a growing branch collides with a branch on another tree that is already present, the growth is rejected and the branch ``retreats'' by a small amount, $0.02\Lambda$, then it deflects by turning in a random orthogonal direction. It then proceeds to lengthen and branch with its new orientation, which could result in more collisions that are similarly deflected. The lengthening and branching processes are repeated to form a highly branched tree; in principle, lengthening and branching could repeat indefinitely, but here the simulation is truncated after a set number of branching events ($B$). Control parameters $\Lambda$ and $\theta$ allow us to test a wide variety of branch geometries.

The goal of these simulations is to investigate the range of geometries that facilitate entanglement via growth. As such, we employ no other mechano- or chemo-sensing behavior; we also do not allow the dendrimers to elastically deform. Any of those behaviors would make entanglement more likely to occur, so we excluded them to keep the focus on geometry. Thus, these simulations can be considered simple random walk models of dendrimer-like growth that investigate the geometries of growing and entangling branches. We also simulated an alternative approach in which we make entanglement even more difficult to achieve; growing tips that collide with existing branches cease growing; these simulations produced results qualitatively similar to what we will detail below (see SI Figure \ref{SF_03}).

\subsection*{Growth assembles configurations that mechanical agitation cannot}
We explored how easily growth entangles trees compared to sequences of translations and rotations. Particularly, we hypothesized that two already-grown and non-entangled trees would have a limit to how close they can be pushed together via translations and rotations. Conversely, we hypothesized that growth could allow branches to penetrate deeper into already-grown trees, resulting in configurations that were irreversible to translations and rotations. To identify these configurations, we measured the proximity of the tree centers of mass after translating and rotating trees together, and after growing trees near each another.

First we grew two trees independently (geometric parameters $\Lambda=4,\theta=45\degree$), and translated one tree towards the other until the branches of the trees collided, as detected by any non-zero intersection of their alpha shapes (Figure \ref{FIG_03}b). After this initial collision, one tree went through a series of small rotations and translations that were mechanically restricted through collision detection. Then, the trees were again pushed together. This process of pushing and agitating was cycled many times. We tracked the distance between the clusters' centers of mass over time (Supplemental Figure \ref{SF_04}), finding that this amount of agitation lead to a plateau in the closest distance the clusters could reach by the 25\textsuperscript{th} cycle. In Figure \ref{FIG_03}d, we plot the histogram of shortest distances achieved by the two clusters. The mean distance achieved was $L_0=96.1\pm 8.8$ simulation units ($N=188$); for comparison, the mean tree diameter is $107.9\pm 0.3$ simulation units. Therefore, the pushing algorithm generally resulted in tree configurations that only weakly penetrated each other's space. We scale all future measurements of tree proximity by the value $L_0$, such that the mean distance achieved for this set of simulations is of unit magnitude.

We next sought to model entanglement from growth. We grew one tree in isolation, and then started growing the second tree $50$ simulation units ($0.52 L_0$) away from the first tree's center of mass (Figure \ref{FIG_03}c). This distance represents about half the mean distance between centers of mass achieved by the agitated trees, and is also located well inside the radius of the first tree. Because the new seedpoint was located slightly inside the radius of the first tree, we checked if there was any initial overlapping volume (which would represent two cells occupying the same space), and generated a new location if there was. This approach resulted in 134 grown configurations. The mean final center of mass separation distance was $55.2\pm 4.6$ units ($0.57 L_0$), substantially closer than through agitation alone ($p\ll 0.001$, $z=8.9$, z-test). The closest pair of agitated trees achieved a center of mass distance of $72.8$ units ($0.76 L_0$); only $1$ out of the total $134$ grown trees was farther apart than that. Further, it is worth pointing out that grown trees achieve this small center of mass distance despite growing randomly in all directions, while agitated trees experience a directional force that is designed to push their centers of mass together. Thus, grown trees appear to readily achieve configurations that are inaccessible via agitation alone.

\subsection*{Growth geometry mediates time needed to entangle}
One of the characteristics of entangling granular materials is that there is a geometric ``sweet spot'' for which entanglement probability is maximized \cite{Gravish2012}. We explored if such a geometric sweet spot also exists in our growing system. We tested many different branch geometries by varying the geometric properties $\Lambda$ (distance between branchpoints) and $\theta$ (angle of the new branches) as shown in Figure \ref{FIG_03}e. In each case, we simulated $100$ different instances, where one tree was grown in isolation, and then a second tree was grown nearby. Then, to quantify entanglement, we dragged the two trees apart along the vector determined by the difference between their centers of mass, and tracked collisions by quantifying the overlapping volume of their alpha shapes. We enumerated the proportion of instances where the two trees collided from this drag experiment. We found that, within the test parameters ranging from $\Lambda=[2,6]$ and $\theta=[15,90]$ and $B=3$ branching events of growth, entanglement was more likely for sparser networks (larger $\Lambda$) and for intermediate branching angles ($\theta=120\degree$) (Figure \ref{FIG_03}f). These results are consistent with previous experiments on entangled granular materials that identified a geometric sweet spot for maximum entanglement probability \cite{Gravish2012}.

We then increased the amount of time the target tree grew, changing the number of branching events, $B$. When trees were grown for a short amount of time, there was little entanglement observed, except for geometries near the sweet spot. At intermediate times, many configurations begin to entangle, but the geometric sweet spot is still easily observable. However, when grown for long enough, even geometries far from the sweet spot begin to entangle, and since the probability of entanglement saturates at $1$, these poorly-entangling geometries ``catch up'' to well-entangling geometries. In Figure \ref{FIG_03}f, we demonstrate this saturation effect as a phase map with 4 panels, the first corresponding to $B=3$ branching events, then $B=4$, $B=5$, and $B=6$. There exist geometries that are not available for entanglement no matter the growth time; these geometries correspond to very dense hyphal networks with no space between the branches (i.e. some seen in Figure \ref{FIG_03}e, and the gray region and line in \ref{FIG_03}f,g). However, for all geometries for which entanglement can occur, the probability of entanglement increases monotonically with time (Figure \ref{FIG_03}g). This phenomenon suggests that for entanglement via growth, the primary role of geometry is not to determine if entanglement occurs, but to determine how much growth is necessary for entanglement to occur. In this sense, the amount of time a branched tree can grow may be more significant than its geometry in determining entanglement.

\section*{Growth ensures tunneling to entangled states}
\begin{figure*}
    \includegraphics[width=\linewidth]{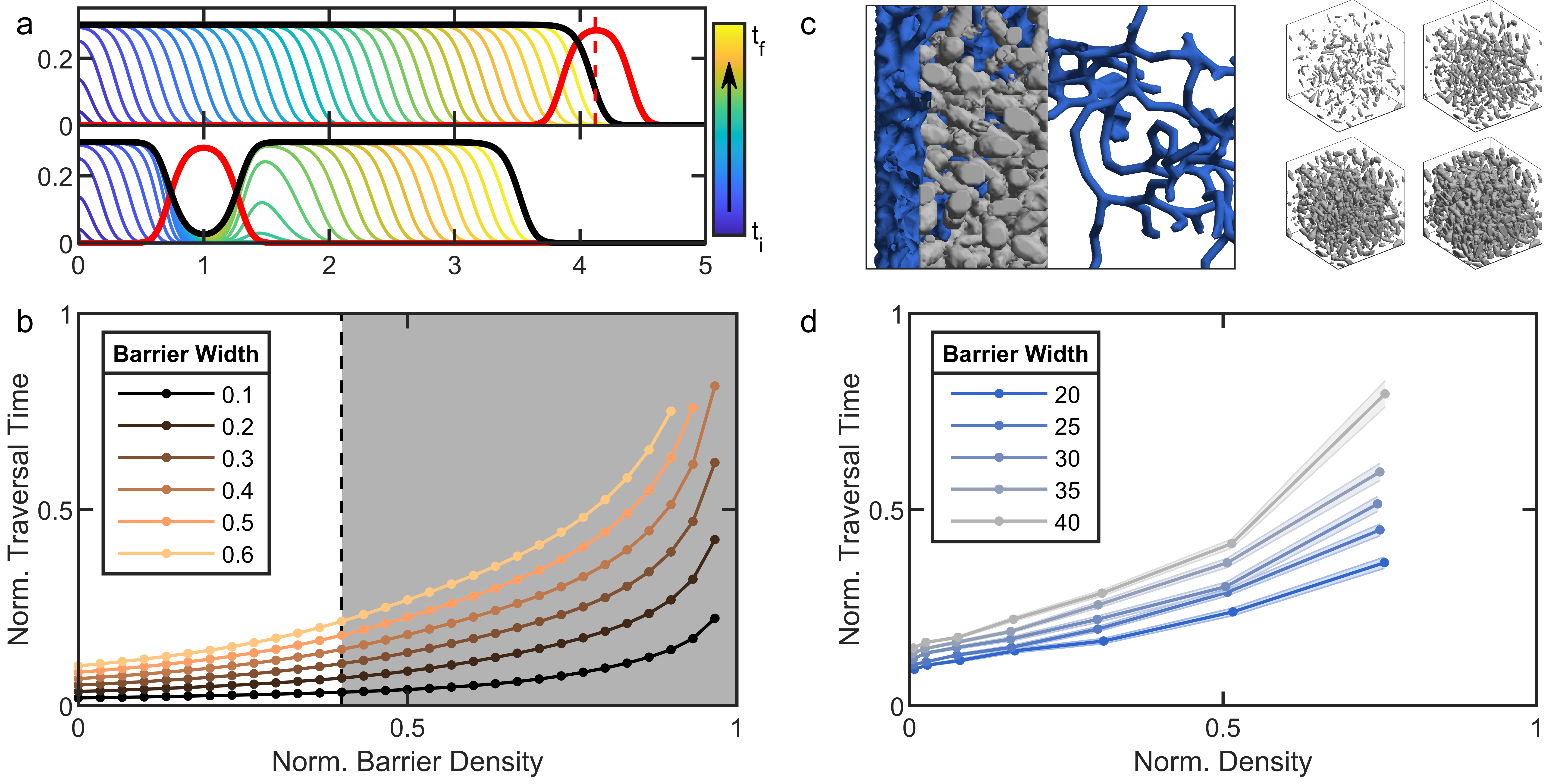}
    \caption{\textbf{Growth ensures tunneling to states unreachable from agitation alone.} \textbf{a} Two numerical solutions to equation (1), plotting $\rho(r)$ vs $r$. Top: Equation 1 is solved in free space, then a barrier (red) is translated towards the grown density field (black) until the two-density sum exceeds $\phi_{max}$ at any location. Dashed line is the closest the barrier reaches. Bottom: Equation 1 is solved when the barrier (red) is present, illustrating tunneling through the barrier. Color bar and arrow indicate direction of proceeding time. \textbf{b} Traversal time of a square barrier vs. the barrier density, from solutions to equation 1 for different barrier widths, normalized by maximum numerical integration time. White region is where the square barrier density is low enough so that the barrier could be pushed completely through the grown density field, illustrating states that are accessible to agitation. Gray region is where the barrier and field cannot be pushed through each other, indicating thermally inaccessible configurations. \textbf{c} Left, Example simulation of traversal of rigid, branched hyphae (blue) through a porous medium (gray). Right, Examples illustrating changing density of the porous medium from 0.02 to 0.23. \textbf{d} Normalized traversal times of  simulations for varying porous medium densities and widths. The x-axis is scaled by the bond percolation threshold for 3d cubic lattices \cite{Sykes1964,Gaunt1983}. Different color lines represent means across 96 different simulations for different barrier widths. For clarity, the standard deviation in traversal times for only one barrier width is displayed.}
    \label{FIG_04}
\end{figure*}

Our results so far suggest that entanglement via growth occurs readily for branched trees that are allowed to grow for a sufficient amount of time, with less dependence on their branching geometry. However, it remains possible that these clusters, and snowflake yeast, are especially ``primed'' for entangling via growth compared to entangling via agitation. We thus sought to test these ideas with an approach that provides maximal leniency for entanglement from agitation, and that lacks a specific geometry.

To do so, we employed a non-geometric, space-filling model. In this model, we do not specify the growth morphology, and do not model contact-based interactions between branches. Instead, we model the density of a branching structure as a spatiotemporal scalar field $\rho(\mathbf{r},t)$. We consider a system with a radially isotropic density, reducing the system to one dimension. The only mechanical rule we impose is that a maximum packing density exists, i.e., the sum of all separate density fields representing different objects is limited by a maximum material packing density, $\sum_i\rho_i(\mathbf{r}) \leq \phi_{max}$, where $0<\phi_{max}\leq 1$. This approach is inspired by simple, but fundamental, physics of close-packed particles and cells cells cannot overlap, and based on their geometry have a maximum packing fraction they cannot exceed \cite{Donev2004, Jacobeen2018a, Day2022}. These packing ``rules'' apply to all real cellular systems, but are also maximally permissive for entanglement via agitation---so long as the sum of two density fields remains less that $\phi_{max}$, they can be pushed together such that they overlap. In fact, for this model it would be possible to push two such objects directly through one another, so long as the packing density at every location remains below the maximum packing density, i.e., $\sum_i\rho_i(\mathbf{r}) \leq \phi_{max}$ remains true everywhere. Clearly, for real, rigid objects, this is not possible. Thus, this model is quite lenient for agitated systems. Nonetheless, we will find that growth easily and inevitably accesses entangled configurations in regimes that are inaccessible to agitation (Figure \ref{FIG_04}).

We begin by modeling the time evolution of a growing system, modeled as a radially isotropic field. Growth can occur in the radial direction, thus increasing the radius, or it can occur in directions that are orthogonal or mis-aligned to the radial vector, thus increasing density in a region of space they already occupy. We model this time evolution as:
\begin{equation}
    \frac{\partial\rho_i}{\partial t} = K \rho_i (1-\frac{\sum_j \rho_j}{\phi_{s}}) + D(r) \nabla^2 \rho_i
\end{equation}
The first term of the right-hand side models the dynamics of increasing density at occupied positions. Once there is material occupying a position $\mathbf{r}$, the density field at this point will increase via growth until it reaches its maximum value, $\phi_s$, with rate of solidification $K$. This logistic term also includes information about other scalar density fields, with which $\rho_i(r)$ must interact. This other material acts as a further cap to the maximum density that $\rho_i$ can reach. The second term in Equation 1 models expansion, i.e., growth into previously unoccupied position $\mathbf{r}$. We model expansion with a diffusion-like second order spatial derivative (with proportionality constant $D(r)$, that varies spatially) due to the stochastic random-walk-like nature of branching events in our simulations. Further, we model the spatial variation of the effective diffusion constant as $D(r)=D_0(1-\frac{\sum_j \rho_j}{\phi_{max}})$ to reflect the slowing rate of expansion when interacting with dense, porous material.

An important characteristic of living, growing materials (such as those in Figure \ref{FIG_01}) is that they often do not grow to fill space, i.e., their grown packing fraction is less than the maximum possible \cite{Jacobeen2018a, Day2022}. Factors such as growth morphology or the uptake and diffusion of nutrients can limit the density to which the organism grows. In our model, we allow for this possibility by explicitly writing the maximum density achieved through the solidification process as $\phi_s$, which may be less than the maximum possible density $\phi_{max}$. We next numerically integrate our partial differential equation model. We will consider two scenarios, representing entanglement via agitation and entanglement via growth. For each scenario, $\phi_s=0.3$, a similar value to experimental measurements of the cellular packing density of snowflake yeast \cite{Day2022}, and the maximum packing density is $\phi_{max}=0.5$.

First, we separately grew two clusters (by integrating Equation 1) until they each reach $\phi_s$ in their center. We then pushed one cluster towards the other, which we refer to as the barrier. Eventually, the cluster reached a position $r_0$ such that $\rho_1(r_0)+\rho_2(r_0)=\phi_{max}$. At this point, the cluster cannot be pushed any farther, as doing so would result in $\phi(r_0) > \phi_{max}$ (Figure \ref{FIG_04}a).

In the second scenario, we grew a barrier until it reaches $\phi_s$ in its center. We then grew a cluster starting a distance $r=1$ away from the center of the barrier, and observed that the cluster grows through the barrier. As the height of the barrier $\sigma_0<\phi_{max}$, growth inevitably tunnels through the barrier to the other side, where it then continues to solidify, entangling the barrier in place (Figure \ref{FIG_04}a). Note that this is a deterministic system; tunneling through the barrier will always happen for these chosen parameters.

Next, we sought to test the impact of the barrier's density and width. We generated step function barriers with varied densities, from $0$ to $\phi_{max}$, and widths, from $w=0$ to $0.6$, to investigate the amount of growth time necessary to traverse the barrier. This traversal time (which we have normalized by the total length of our numerical simulations, i.e., the total integration time) diverges as the barrier height approaches the maximum density. Conversely, traversal time does not diverge with barrier width, implying that, in principle, even a very wide barrier will eventually be traversed. The gray region of Figure \ref{FIG_04}b illustrates a regime where pushing the cluster all the way through the barrier is impossible, because the sum of the two density fields would together exceed $\phi_{max}$. Growing fields can traverse the barrier even in this gray zone because growth can proceed without exceeding $\phi_{max}$. This means that, even in the case of this model, which is quite lenient to translating fields directly through one another, growth still accesses configurations that translation cannot achieve. However, traversal times within the gray region can be quite low (Figure \ref{FIG_04}b).

We next sought to test some of these predictions via simulations of dynamic, growing hyphae in three dimensions. We use a sample cube of SEM data from snowflake yeast experiments to generate a porous barrier. The density of this block is controlled by eroding or dilating the voxels of the 3d data sample (see Methods and Fig. \ref{FIG_04}c). Then, we employ the branched-tree growth simulation used in Figure \ref{FIG_03} to explore paths through the porous block. We measure the traversal time when any branch of the growing tree reaches the opposite side of the porous block. In Figure \ref{FIG_04}, we show one simulation, where the branch network (blue) starts on the left side of the porous block, and then grows, eventually traversing a path through the gray porous block to the other side. We tracked the traversal time for $96$ simulations of each barrier density and width (Figure \ref{FIG_04}d). We found that, in qualitative agreement with the mean-field model, the traversal time increases superlinearly for increasing density.

\section*{Experimental tests of growing entanglement}
\begin{figure*}
    \includegraphics[width=\linewidth]{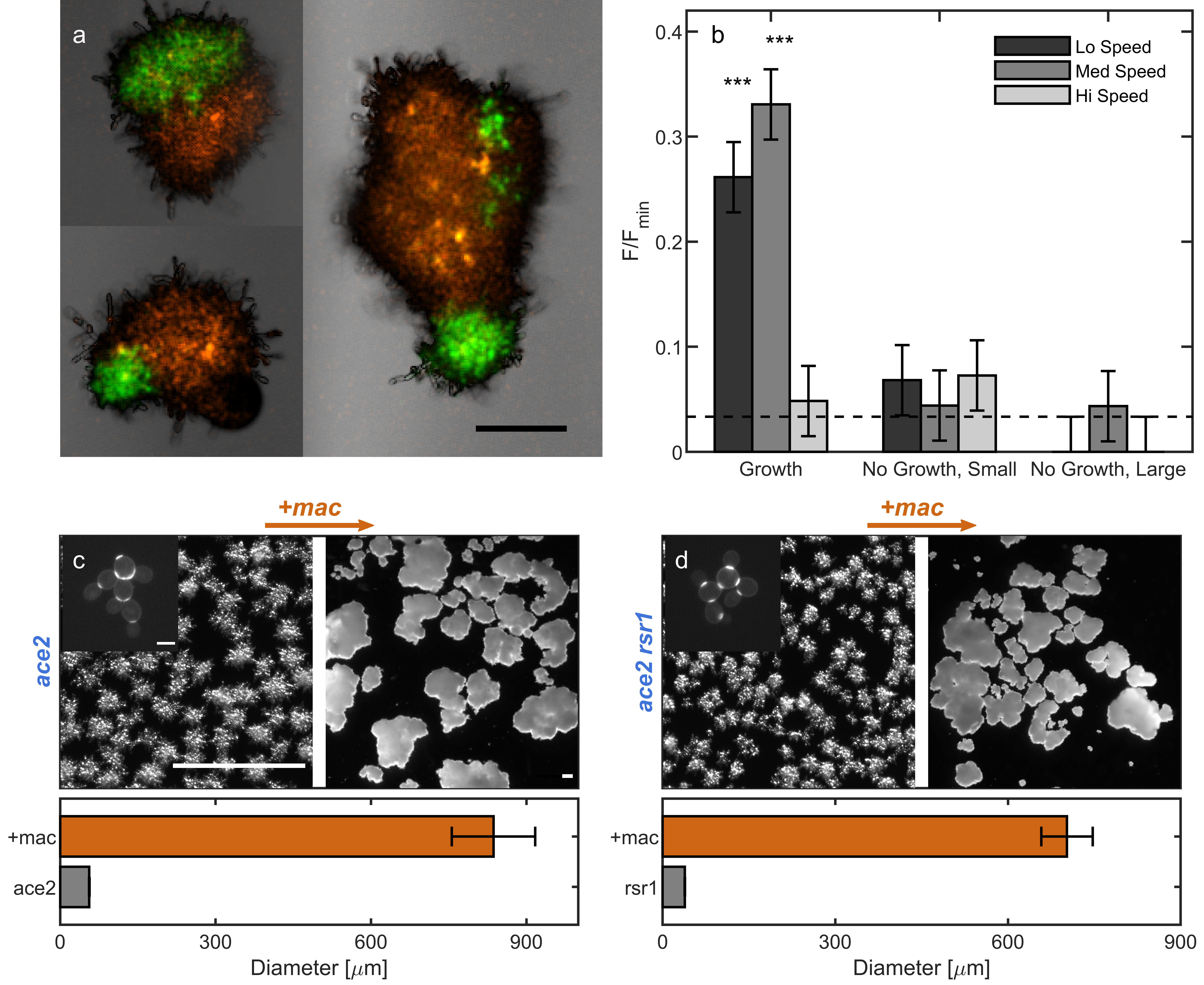}
    \caption{\textbf{Growing branches entangle readily}. \textbf{a}, 3 examples of combination-colored, entangled yeast clusters. Scale bar is \SI{100}{\micro\meter}. \textbf{b}, Proportion of clusters observed to be entangled for various treatment types, normalized by the proportion of the minimum color type. Horizontal dashed line is a measurement of algorithm error via a control where no entanglement is expected (i.e. immediately pipetting a mix of green and red clusters onto a slide without any agitation or growth). Error bars are also drawn from this empirical measurement of algorithm precision. Three stars indicates $p<0.001$ significance level. \textbf{c} Top Left: $ace2$ snowflake yeast that does not entangle. Inset: higher-magnification image of $ace2$ stained with calcafluor white that brightly highlights bud scars to show the characteristic snowflake yeast pattern. Inset scale bar is \SI{5}{\micro\meter}. Top Right: we apply genetic mutation set 1 ($mac = clb2 + cln3 + gin4$), inducing changes that lead to entanglement. The scale bars in Top Left and Top right are both \SI{300}{\micro\meter}. Bottom: Mean diameter (and standard error) of these two strains. \textbf{d} The same information as \textbf{c}, but for the mutant line $ace2 rsr1$. The images have the same scale as their counterparts in \textbf{c}.}
    \label{FIG_05}
\end{figure*}

Finally, we seek to experimentally test the idea that entanglement via growth depends heavily on time scales. To do so, we use the snowflake yeast model system of undifferentiated multicellularity, which has recently been shown to evolve branch entanglement as a mechanism of generating increased multicellular toughness \cite{Ratcliff2012, Bozdag2023}.

\subsection{Altering timescales to encourage entanglement between clusters}
One of the predictions of our entanglement models from above is that branched trees can easily grow into entangled configurations so long as they remain near each other and grow for long enough. We sought to test this prediction experimentally by growing, agitating, and imaging populations of differentially-labeled (red and green) snowflake yeast. It was previously demonstrated that separate clusters do not entangle when grown in a shaking incubator at 225 rpm \cite{Bozdag2023}. But, based on the above simulations, we hypothesized that separate clusters will entangle if shaken at lower speeds, as they will spend more time in contact with each other.

To test this hypothesis, we broke groups of red- and green-fluorescent snowflake yeast clusters into small pieces by compressing them between glass slides, and then grew the red and green pieces in a single culture tube. After incubation, we vortex-mixed each tube to ensure that any observed entanglements were mechanically stable, and then imaged clusters to determine if distinct red and green trees became entangled. We determined the experimental error via an empirical control where we expect no entanglement to occur; we cultured red- and green-fluorescent strains in separate tubes overnight, then mixed them in a single vial and imaged them immediately. All error bars in Figure \ref{FIG_05} are from this control experiment.

When incubated in growth medium at low and medium shaking speeds (50 and 150rpm), entanglement between distinct red and green trees readily occurred ($26\%$, $N=273$ and $33\%$, $N=299$ of examples Figure \ref{FIG_05}a), quantified by the relative proportions of combination-colored trees compared to the total proportion of red and green trees (Figure \ref{FIG_05}b, see Methods). At high shaking speeds (250 rpm), entanglement was rare. This stark difference occurs because at 50 rpm and 150 rpm, yeast groups remain settled near the bottom of the tube, presumably interacting with the same neighboring trees for multiple rounds of cell division (see SI movie 1). At 250 rpm, clusters are dispersed throughout the fluid, and therefore pairs of clusters do not stay near each other for sufficient times to grow entangled. 

To directly test the effect of growth compared to mechanical agitation alone, we incubated some samples in a saline solution that inhibits cell division but keeps cells alive. In the experiments with growth media, snowflake yeast clusters start small but grow to large sizes. We thus performed controls in saline solution for both small clusters and large clusters, with size distributions matching the start and end points of the growth experiment, i.e., we performed experiments with both small (broken) trees and large (unbroken) trees. In all cases where we observed entanglement via growth (i.e. low and medium shaking speeds with growth medium), the proportion of entangled red-green trees was significantly higher in culture tubes with growth than in those without growth ($p<0.001$, z-test), suggesting that random translations and rotations of the yeast branches due to agitation is not sufficient to entangle separate clusters. 

\subsection{Genetically altering branch geometries}
The above work suggests that the phenomenology of entanglement via growth and entanglement of non-living materials are qualitatively different. On the one hand, non-living materials entangle only if they are situated near the geometric ``sweet spot'' \cite{Gravish2012}. On the other hand, our simulations (Figures \ref{FIG_03} and \ref{FIG_04}) predict that entanglement via growth can occur even for branching geometries that are far from the ``sweet spot'' if the organisms are given enough time to grow. In other words, non-living materials only entangle with ideal geometries, while we predict that entanglement via growth can occur even with geometries that are far from ideal. In this section, we will test these ideas experimentally.

To do so, we genetically engineered two different strains of microscopic snowflake yeast, each with a different budding geometry. The first mutant is created by knocking out the gene $ace2$ in single celled yeast; this is the ``ancestral'' strain used for the multicellularity long-term evolution experiment (MuLTEE) \cite{Bozdag2023}. This strain of snowflake yeast tends to produce distally polar buds with fairly regular polar angles of $\langle\theta\rangle=34\pm17\degree$ (Supplemental Figure \ref{SF_05}). The second mutant is created by knocking out the genes $rsr1$ and $ace2$. Previous works have shown that knocking out the gene $rsr1$ causes buds to appear in random locations on the yeast cell surface \cite{Casamayor2002}. In our engineered mutants, the mean budding angle was $\langle\theta\rangle=57\pm31\degree$ (Supplemental Figure \ref{SF_05}) which is a much broader distribution than in the $ace2$ snowflakes ($p=0.005$, $t=2.9$, $df=40$, two-sample t-test; see Supplemental Figure \ref{SF_05}). Importantly, neither of these mutants entangle; mechanical stresses cause intercellular bonds to fracture, splitting the organism into separate pieces \cite{Ratcliff2015, Jacobeen2018a}.



We next will determine if clusters with these different budding angle geometries can entangle. We previously demonstrated that a mutation set ($clb2, cln3, gin4$), here called $mac$, can cause the $ace2$ mutant to entangle (we repeated these measurements anew here, Figure \ref{FIG_05}c\&d) \cite{Bozdag2023}. When we created this mutant strain of snowflake yeast with highly elongated cells and stronger intercellular bonds, mean group diameter increased by a factor of $14.9$ (Figure \ref{FIG_05}c\&d, $p\ll0.001$, two-sided t-test, $df=600, t=-33.5$), which corresponds to a change in volume of $>1000$-fold. In prior work, we have shown that the onset of entanglement leads to a similar increase in group volume \cite{Bozdag2023}. Therefore, we will use such an increase in group size as a proxy for entanglement. If entanglement via growth requires budding angle geometry to be near a geometric sweet spot, then a snowflake yeast mutant with $rsr1$ and $mac$ (i.e. $ace2+rsr1+clb2+cln3+gin4$) should not entangle, since the budding angle distribution is quite different and more spread. However, if adding the $mac$ mutations to an $rsr1$ mutant does result in entanglement, this would experimentally demonstrate that entanglement via growth can occur for a wide range of geometries.


We next constructed the $ace2+rsr1+mac$ mutant and measured its size. In agreement with our predictions from simulations, we find that $rsr1+mac$ mutants do in fact entangle, increasing mean group diameter by a factor of 18.3 (Figure \ref{FIG_05}c\&d, $p\ll0.001$, $t=60.6$, $df=1053$, two-sided t-test). This test therefore confirms that the original snowflake yeast budding geometry is not necessary for entanglement to proceed. Instead, entanglement via growth can occur for various budding geometries, including those that have been randomized, and are potentially far from any sweet spot.

\section*{Discussion}
Here, we used a combination of experiments, simulations, and theory to show that growth of branching, rigid trees more readily leads to entanglement than agitation alone. We argued for and experimentally found that growth can produce effectively inaccessible configurations, i.e., ones that are impossible to disassemble. In simulations and experiments, we showed that branching growth readily accesses these configurations, even without higher-evolved sensing behaviors. Finally, while geometric properties such as branch diameter clearly play a role in the frequency and strength of entanglements, we showed through numerical methods that, given the right conditions for entanglement to occur, growth inexorably tunnels into configurations that are impossible to access via agitation alone. Combined, this evidence supports the idea that entanglement in growing systems is relatively easy to achieve, and more dependent on timescales than geometry. 

There are two ways that non-living systems are known to entangle. First, they can mechanically agitate separate pieces into configurations where they wrap around each other \cite{Brown2012, Gravish2012, Meng2018, Ozkan-aydin2021, Tuazon2022}. Second, entanglement can be triggered via synthesis of new bonds between previously separate chains \cite{Tang2014, Edwards2020, Kim2021, Norioka2021}. We have not explicitly compared entanglement via growth to the latter case. But it is worth noting that entanglement triggered through new bond formation is a carefully engineered process. So far, such processes have been studied with polymers composed of modular, alternating blocks of coils and elastin-like domains \cite{Tang2014, Edwards2020}, so that new bonds are selectively triggered in particular locations along the polymer. Otherwise, cross-linking of chains becomes more frequent than entanglement events, and the polymer gel loses its entanglement-derived qualities \cite{Kim2021}. This kind of precise engineering is currently much more difficult to achieve with living systems. While extant complex multicellular organisms may be capable of the precision needed for selectively cross-linking entanglements, entanglement through growth is likely a more relevant mechanism for establishing entanglements in simple or nascent multicellular groups.

It was recently demonstrated that even in cross-linked gels, entanglement is responsible for increased toughness \cite{Tang2014,Kim2021, Edwards2020}. Material toughness is an important property for many organisms and organism collectives, especially those that need to avoid fracture. For instance, toughness is an important characteristic of cartilage \cite{Jackson2022} and other collagen networks \cite{Burla2020}, especially as joint degeneration progresses with age or injury \cite{Decker2017}. Animal collectives have also been shown to actively entangle by bending limbs \cite{Ozkan-aydin2021, Hu2016} as a mechanism for holding themselves together under external stresses like shear flows. Therefore, even in living systems where cross-linkers are known and studied, accounting for entanglement may be important.

In this paper, we have shown that entanglement is a common and robust phenomenon in living systems that grow as branching trees with permanent cell-cell bonds. Such bonds are a frequent evolutionary outcome in the transition to multicellularity, as exemplified by fungi, plants, red, green, and brown algae, and filamentous bacteria \cite{Day2022a}. We have demonstrated that entanglement via growth does not depend on specific geometries or morphologies, but rather on the time scales of growth and interactions. Indeed, within the snowflake yeast model system, entanglement evolves within just 3,000 generations of selection for larger size \cite{Bozdag2023} - a geological blink of an eye. Rather than requiring substantial developmental innovation, we suggest that entanglement may be one of the first mechanisms evolved by branching multicellular organisms under selection to grow tough bodies capable of withstanding internal strains from cell division, or external stresses from the environment. Despite the ease with which entanglement can evolve, and its convergent evolution across many multicellular clades, much remains to be discovered about the role of entanglement as a mechanism for generating tough, strong, multicellular materials.

\begin{acknowledgments}

\end{acknowledgments}

\appendix
\section{Culturing and sample preparation}
Multicellular yeast groups were sampled from an ongoing long-term evolution experiment (MuLTEE \cite{Bozdag2023}). These anaerobic multicellular yeast clusters were evolved from an ancestral multicellular ``snowflake'' petite yeast without a functional copy of the gene \textit{ace2}. When the ace2 gene is not expressed, the final stage of cell division is not completed, and mother-daughter cells remain attached at the chitinous bud site. Since all cells are attached directly to their mothers, snowflake groups form a fractal-like branched tree collective.

Yeast was generally cultured in \SI{10}{\milli\liter} YEPD media (\SI{10}{\gram} yeast extract, \SI{20}{\gram} peptone, \SI{20}{\gram} dextrose for \SI{1}{\liter} DI water). To keep yeast alive yet prevent further growth, we used a saline solution of $0.85\%$ sodium chloride dissolved in DI water. Glass culture tubes were then cultured overnight at 30C in a variable-speed shaking incubator (Symphony Incubating Orbital Shaker model 3500I). We selected 50rpm, 150rpm, and 250rpm shaking speeds to vary the agitation strength. 

\subsection*{Mechanical agitation and population size measurements}
To test if mechanical agitation could disassemble grown yeast clusters, yeast clusters were first grown in overnight culture, then agitated, and then imaged to obtain a population-level size distriubtion. After culturing, we used wide-tip \SI{1000}{\micro\liter} tips to pipette \SI{1}{\milli\liter} of culture (making sure to gently shake first) into two different microcentrifuge tubes. Each tube was vortexed once at medium vortex speed (5, \textit{model of vortexer}) for 5 seconds. Then, one tube was left without any more vortexing, and the other was vortexed for an additional 5 seconds at a stronger vortexing speed (7). \SI{100}{\micro\liter} was sampled from each tube into fluorodishes so that the clusters were not flattened by a microscope slide, and imaged under brightfield using a Zeiss Axio Zoom V16 microscope.

After imaging, custom MatLab scripts segmented and binarized the clusters from the background. These scripts use a combination of watershedding, morphological segmentation, and filtering to separate proximate clusters (code attached). Cluster cross-sectional area was measured, and used to estimate cluster diameter using a spherical approximation.

\subsection*{Fluorescent tagging}
To visualize entanglements between different groups of snowflake yeast, we isolated a single snowflake genotype from PA2, t600 (strain GOB1413-600), and engineered it to constitutively express either green or red fluorescent proteins. To do that, we amplified the prTEF-GFP-NATMX construct from a pFA6a-eGFP plasmid and the prTEF-dTOMATO-NATMX construct from a pFA6a-tdTomato plasmid. We then separately replaced the URA3 open reading frame with GFP or dTOMATO constructs in an isogenic single strain isolate following the LiAc transformation protocol \cite{Gietz2007}. We selected transformants on Nourseothricin Sulfate (Gold Biotechnology Inc., U.S.) YEPD plates and confirmed green or red fluorescent protein activity of transformed macroscopic clusters by visualizing them under a Nikon Eclipse Ti inverted microscope.

\subsection*{Genetic manipulation of geometry}
To manipulate snowflake geometry, gene deletions were performed using standard PCR-product based yeast transformation tecnniques\cite{gietz2007quick}. Plasmid pYM25 bearing the hphNT1 gene for hygromycin resistance, pYM42 bearing natNT2 for nourseothricin resistance, and plasmid pYM27 bearing kanMX4 for G418 resistance were used as PCR templates\cite{janke2004versatile} for the deletion of genes using oligonucleotides with 50 base pairs of flanking sequence from around each reading frame. Each gene was deleted individually in the Y55 homozygous diploid background previously used in the Ratcliff laboratory\cite{ratcliff2015origins} or the multicellular GOB8 \textit{ace2$\Delta$:KANMX / ace2$\Delta$::KANMX} strain created from the same background. Random budding small clusters (Strain X, genotype \textit{ace2$\Delta$::KANMX / ace2$\Delta$::KANMX, rsr1$\Delta$::natNT2 / rsr1$\Delta$::natNT2}), quadruple mutants (strain AJB770, genotype \textit{ace2$\Delta$::KANMX / ace2$\Delta$::KANMX, cln3$\Delta$::hphNTI / cln3$\Delta$::hphNTI, clb2$\Delta$::kanMX4 / clb2$\Delta$::kanMX4, gin4$\Delta$::natNT2 / gin4$\Delta$::natNT2}) and quintuple mutants (strain AJB799, genotype \textit{ace2$\Delta$::kanMX4 / ace2$\Delta$::kanMX4, cln3$\Delta$::hphNTI / cln3$\Delta$::hphNTI, clb2$\Delta$::kanMX4 / clb2$\Delta$::kanMX4, gin4$\Delta$::hphNTI / gin4$\Delta$::hphNTI, rsr1$\Delta$::natNT2 / rsr1$\Delta$::natNT2}) were constructed by repeated sporulation and mating of these single-deletion strains.

\subsection*{Mixed culture preparation}
First, two separate tubes of green and red fluorescent yeast were grown overnight. Then, \SI{150}{\micro\liter} was sampled from each tube. The sample was spun down in  a centrifuge, and the YEPD supernatant was removed via pipetting. \SI{300}{\micro\liter} DI water was added, and the spin-down-rinse cycle was repeated. \SI{100}{\micro\liter} was immediately transferred to a tube containing \SI{10}{\milli\liter} saline solution; these were the large controls. The remaining \SI{200}{\micro\liter} was centrifuged and most water was pipetted away. The remaining paste was transferred onto a sterilized glass microscope slide. Another slide was placed on top, and fingertip pressure and shear was added to break the snowflake yeasts into small pieces. The crushed paste was rinsed with sterile DI water into a microcentrifuge tube. \SI{100}{\micro\liter} was transferred each into culture tubes containing YEPD and saline solution, forming the growing sample and the small control, respectively. All three culture tubes were then placed in the shaking incubator overnight. This process was repeated anew for each shaking speed tested.

\section{Confocal microscopy}
To make a population-level measurement of entanglement likelihood, we took population-level images via confocal microscopy. First, \SI{100}{\micro\liter} was sampled from each culture tube, and vortexed to ensure observation of only strong entanglements. The sample was then pipetted onto microscope slides with a shallow, round depression so that clusters were not crushed by the microscope coverslips. The population was then imaged using a confocal microscope (Nikon A1R) at a few different z-levels to ensure that any amount of fluorescence was captured. The z-stacks were later compressed into a maximum intensity projection for each color channel.

\subsection*{Counting red, green, and tangled clusters}
Transmission field images were segmented using custom MatLab scripts. Within each segmented region, color channels were binarized. We could therefore count the number of pixels considered red, and the number considered green, for each segmented region. If the fraction of pixels colored red exceeded a threshold value, the cluster was labeled red. Simultaneously, if the fraction of pixels colored green exceeded a threshold value, the cluster was labeled green. If neither red nor green pixel fractions exceeded the threshold value, then the cluster was labeled as unknown, and was discounted from further analysis. If both red and green pixel fractions exceeded the threshold value, the cluster was labeled as both red and green, and considered entangled. For each experiment, this threshold value was tuned to maximize image analysis efficiency. After preliminary image analysis, clusters that were labeled as entangled were visually checked for accuracy.

To gauge the population-level fraction of all clusters that were entangled, we counted the number of clusters labeled each type. Out of all clusters imaged from a population $N$, there were some labeled entirely red $N_r$, some labeled entirely green $N_g$, and some labeled both red and green $N_e$, such that $N=N_r+N_g+N_e$. Based on initial concentrations and growth rates, it was possible that for different experiments, the initial number of red and green clusters in one tube was different. To account for this difference, the fraction of all entangled clusters, $N_e/N$, was normalized by the smaller value of either $N_r/N$ or $N_g/N$ for each shaking speed, as in $n_e=N_e/N_r$. This is the measurement reported in Figure 2.

Because the counting analysis method fundamentally relies on cluster proximity, it is possible that two separate, nonentangled clusters, one red and one green, happen to locate next to one another for imaging, such that the analysis algorithm counts the pair as entangled even though they are not. To measure this experimental imprecision, we ran a control experiment where we do not expect entanglement to occur. Separated red and green clusters were pipetted into the same microcentrifuge tube, and then immediately imaged without allowing time for growth. The measured value for $n_e$ for this experiment was $0.033$, and was taken as the experimental error for the remaining experiments.

\section{Scanning Electron Microscopy}
\begin{figure}
    \centering
    \includegraphics[width=\linewidth]{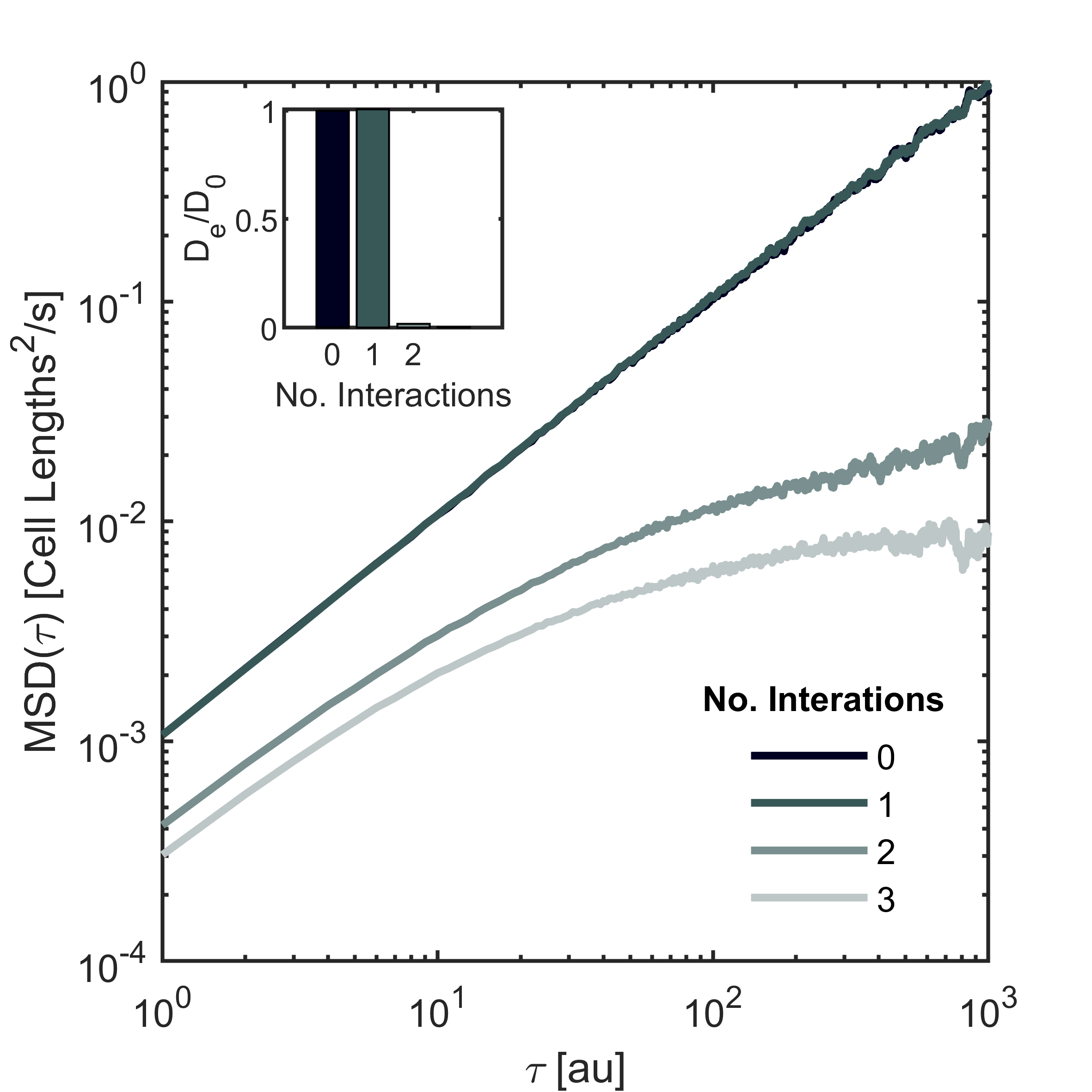}
    \caption{Mean squared displacement plots for four circumstances where one branch (yellow, Figure \ref{FIG_02}a.iii) is agitated with respect to others, with 0 interactions, 1 interacting branch, 2 interacting branches, and 3 interacting branches.}
    \label{SF_01}
\end{figure}

Since yeast cells have thick cell walls that limit the effectiveness of optical microscopy, we used scanning electron microscopy to obtain 3-dimensional structural information. We used data from the same experiments reported in \cite{Day2022} and \cite{Bozdag2023}. In those experiments, we used a Zeiss Sigma VP 3View scanning electron microscope (SEM) equipped with a Gatan 3View SBF microtome installed inside a Gemini SEM column to obtain high resolution images of the internal structure of snowflake yeast groups and locate the positions of all cells. All SEM images were obtained in collaboration with the University of Illinois’s Materials Research Laboratory at the Grainger College of Engineering. Snowflake yeast clusters were grown overnight in YPD media, then fixed, stained with osmium tetroxide, and embedded in resin in an eppendorf tube. A cube of resin \SI{200}{\micro\meter} x \SI{200}{\micro\meter} x \SI{200}{\micro\meter} (with an isotropic distribution of yeast clusters) was cut out of the resin block for imaging. The top surface of the cube was scanned by the SEM to acquire an image with resolution $\SI{50}{\nano\meter}$ per pixel (4000 × 4000 pixels). Then, a microtome shaved a 50-nanometer-thick layer from the top of the specimen, and the new top surface was scanned. This process was repeated until 4000 images were obtained so that the data cube had equal resolution in x, y, and z dimensions.

The resulting voxel representation of the interior of one cluster was then binarized and segmented using custom Python scripts. Connected cells were identified using the nearest neighbor algorithm. From a particular sub-volume of the data cube, we identified 38 connected components, which we will call branches, that were not connected to one another except through mechanical tangling events. Surface data of each branch was obtained by using the surface mesh tool in Mathematica 12, as previously described in \cite{Bozdag2023}. New to this study, the surfaces were imported into Blender and remeshed using the Blender remesh tool to lower the total number of datapoints on the surface. This allowed for faster computation speeds. Then, alpha shapes of these branches were created. Alpha shapes are a generalization of the convex-hull method that allows for nonconvex shapes. The alpha value can be tuned to allow for more or less concavity of the shape. But further, there are fast algorithms for computing intersections of alpha shapes, which was desirable here.

\section{3D structural construction and manipulations}
We chose several individual branches from our list to perform computational manipulations as described in the main text. These manipulations involved either translations or rotations of all points representing the branch, which were carried out via matrix multiplication schemes. All of these manipulation algorithms were written as custom functions in MatLab.

Random rotations and translations were constructed as follows. Random translation steps were sampled from a uniform distribution on the domain $[-.5,.5]$, for each coordinate $x$, $y$ and $z$. Then, the resulting vector was normalized to have length $a$, where $a$ is the step size input. Random rotations were created by first choosing a random axis, then rotating by a chosen angle magnitude around this axis. The rotation axis was randomly chosen by selecting a random number from a uniform distribution on the domain $[-.5,.5]$ for each coordinate $x$ $y$ and $z$, and then normalized to have unit magnitude.

During our mechanical agitation procedures, overlapping volumes were calculated using MatLab's built-in alpha shape capabilities, which allows for fast and accurate computations.

\subsection*{Choosing a threshold for the point of first contact}
For drag experiments, we chose a threshold overlapping volume of 1 cubic micron that marked the ``point of first contact''. We chose this value because it was close to a value for which the force exerted on each cell was $1/10$ the magnitude of force previously measured to fracture bonds between snowflake yeast cells \cite{Jacobeen2018a}. Following a Hertzian model of an elastic material, the force exerted by overlapping two elastic spheres of equal radius $r$ by a distance $d$ is $F=4/3 Y \sqrt{(r/2)}d^{3/2}$, where $Y$ is Young's modulus. The volume of overlap between the two spheres is $V=\pi/12(2r-d)d^2$. Inputting a force of $0.05$\unit{\micro N}, a Young's modulus of $1$\unit{\mega Pa}, and an effective radius of $5$\unit{\micro\meter}, we solve for the appropriate deflection and find an overlapping volume of $0.05$ cubic microns. Multiplied by 20 cells, about the number of cells in the entangled branches, returns a net overlapping volume value of 1 cubic micron.

\subsection*{Measuring bud scar angle distributions}
Bud scar angle distributions were measured for 2 strains: ancestral, $ace2$ knockout yeast cells, and $ace2 rsr1$ genetic mutants. Clusters were cultured overnight in separate YEPD tubes at $30C$ and $250$rpm shaking speed. The next day, bud scars (which are rich in chitin) were stained with calcafluor white. To stain the cells, we took a \SI{500}{\micro\liter} sample of each tube, then centrifuged them and removed the supernatant, and resuspended the sample in sterile DI water. We repeated this process once more, but this time, did not resuspend the pellet. Then, \SI{15}{\micro\liter} of $\SI{1}{\milli\gram}$/\SI{}{\milli\liter} calcafluor stock was diluted in \SI{500}{\micro\liter} 1x Phosphate buffer solution (PBS). \SI{250}{\micro\liter} of the prepared calcafluor solution was added on top of the two pellets. After vortexing, we incubated the tubes in darkness at room temperature for 25 minutes. The tubes were centrifuged, the pellets removed, and the cells were resuspended in sterile DI water. Then, \SI{30}{\micro\liter} was sampled from each tube onto separate glass slides and taken for imaging on a Nikon A1R confocal microscope with a 60x oil objective. We used a \SI{450}{\nano\meter} wavelength laser to excite the calcafluor stain, and took z-stacks of cells to obtain 3-dimensional bud scar information. We examined 9 GOB21 ancestor cells and 12 rsr1 mutant cells, each with 2 bud scars in addition to their one birth scar. Custom Fiji and MatLab scripts segmented the cells, fit an ellipsoid of revolution to each, segmented the bud scars (which fluoresce very brightly), and measured their polar orientation with respect to a birth scar that is always located at the south pole of the cell. The polar angle measured here is $0\degree$ at the north pole, $90\degree$ at the equator, and $180\degree$ at the south pole. The distribution of polar angles is reported in Supplemental Figure \ref{SF_05}.

\begin{figure}
    \includegraphics[width=\linewidth]{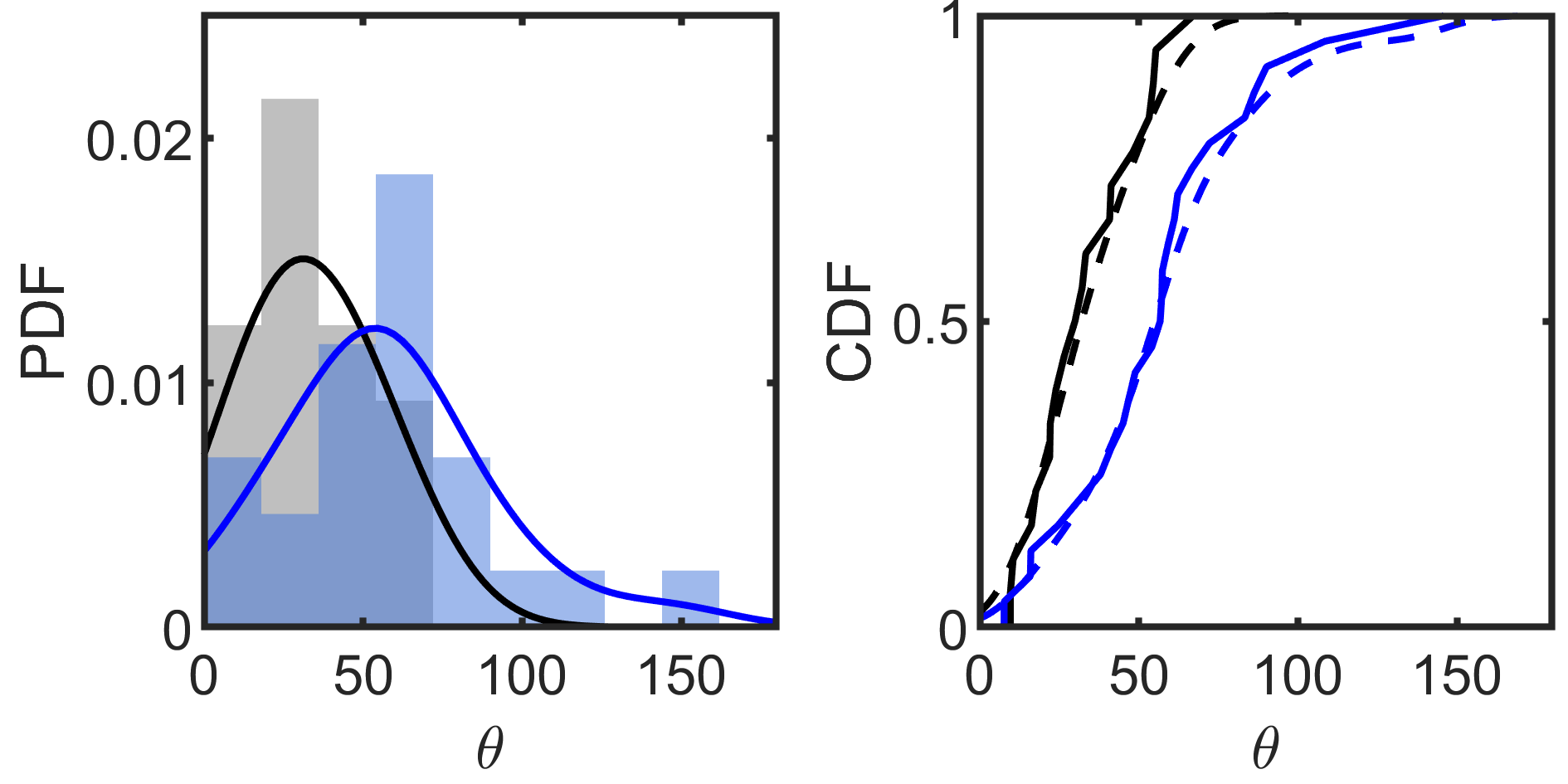}
    \caption{Polar angle distribution for GOB21 and RSR1 yeast cells. \textbf{Left} PDFs for polar angle for ace2 (gray) and $ace2 rsr1$ (blue). Solid lines are kernel smoothed estimates of the density distribution. \textbf{Right} CDFs of the two distributions. Solid lines are the empirical cdf, dashed lines are kernel-smoothed estimates. The two distributions are statistically different via a t-test with $t=2.94, p=0.005, df=40$.}
    \label{SF_05}
\end{figure}

\subsection*{Measuring group size}
To measure group size of each of the four strains of yeast clusters, groups were grown overnight in YEPD at $30C$ and $250$rpm shaking speed. Then, \SI{500}{\micro\liter} were sampled from each tube into a fluorodish so that clusters were not broken into pieces by a glass coverslip. We then imaged the clusters on a Zeiss Axio Zoom V16 microscope. Custom scripts segmented the clusters from these images and measured their cross-sectional area. This area was used to calculate an effective diameter via $d=2*\sqrt{A/\pi}$.

\subsection*{Measuring single cell size}
To measure the single cell size of ancestral $ace2$ yeast cells and randomly-budding $rsr1$ cells, yeast clusters were imaged on a Nikon widefield inverted microscope with a 40x objective. 10 individual cells were segmented from these images, and their maximum diameter was measured using the Fiji length measurement tool. The ancestral snowflake yeast cells had a mean measured diameter of $\SI{7.0\pm0.8}{\micro\meter}$, and the randomly budding mutant had measured mean diameter $\SI{6.6\pm0.3}{\micro\meter}$. A two-sample t-test returns $t=1.54$, $df=18$, and $p=0.14$.

\section{Branched tree simulations}
\subsection*{Growing trees}
We created custom simulations for growing, branching, dendrimer-like objects in 3 dimensions in MatLab. We call these objects trees. Each tree is created by starting with 6 seedpoints, which represent the hyphal tip. Each hyphal tip possesses three properties: a location, an orientation, and the number of steps that have occurred since it last split. We continually track the locations and orientations of each hyphal tip; additionally, we add these locations to a growing list of tip locations for all previous timepoints. Each timepoint, all hyphal tips walk forward one unit in the direction they are oriented; their locations are then updated. A disk of a selected radius is constructed around the hyphal tip such that the plane of the disk is orthogonal to tip's orientation. An integer number of evenly spread points within this disk are then added to the list of all locations previously occupied by the tree.

To start, all seed directions were the 6 cardinal directions (-x and +x, -y and +y, and -z and +z). The hyphal tips then extended until they reached a threshold number of steps from their starting location. At this point, each hyphal tip splits into two tips, each occupying the same location but with different orientations. The angle between the two orientation vectors was varied between $[30,180]\degree$ for our simulations. The azimuthal orientation was randomly selected from a uniform distribution on the domain $\phi=[0,2\pi)$. After splitting, the algorithm continues to track a (now larger) list of the hyphal tips. This process continues for a set number of iterations.

\subsection*{Deflection of branches}
When branches of the same tree encounter one another, they do not interact. When branches of one tree encounter a different tree, they deflect. Collisions were detected by computing the overlapping volume of the alpha shapes of the two trees. If there was any overlap, it was identified. Then, the hyphal tip that penetrated the alpha shape of the second tree retreated a small amount (0.5 units backwards). It then changed its orientation by randomly selecting an orientation orthogonal to its last orientation. 

\begin{figure*}
    \includegraphics[width=.8\linewidth]{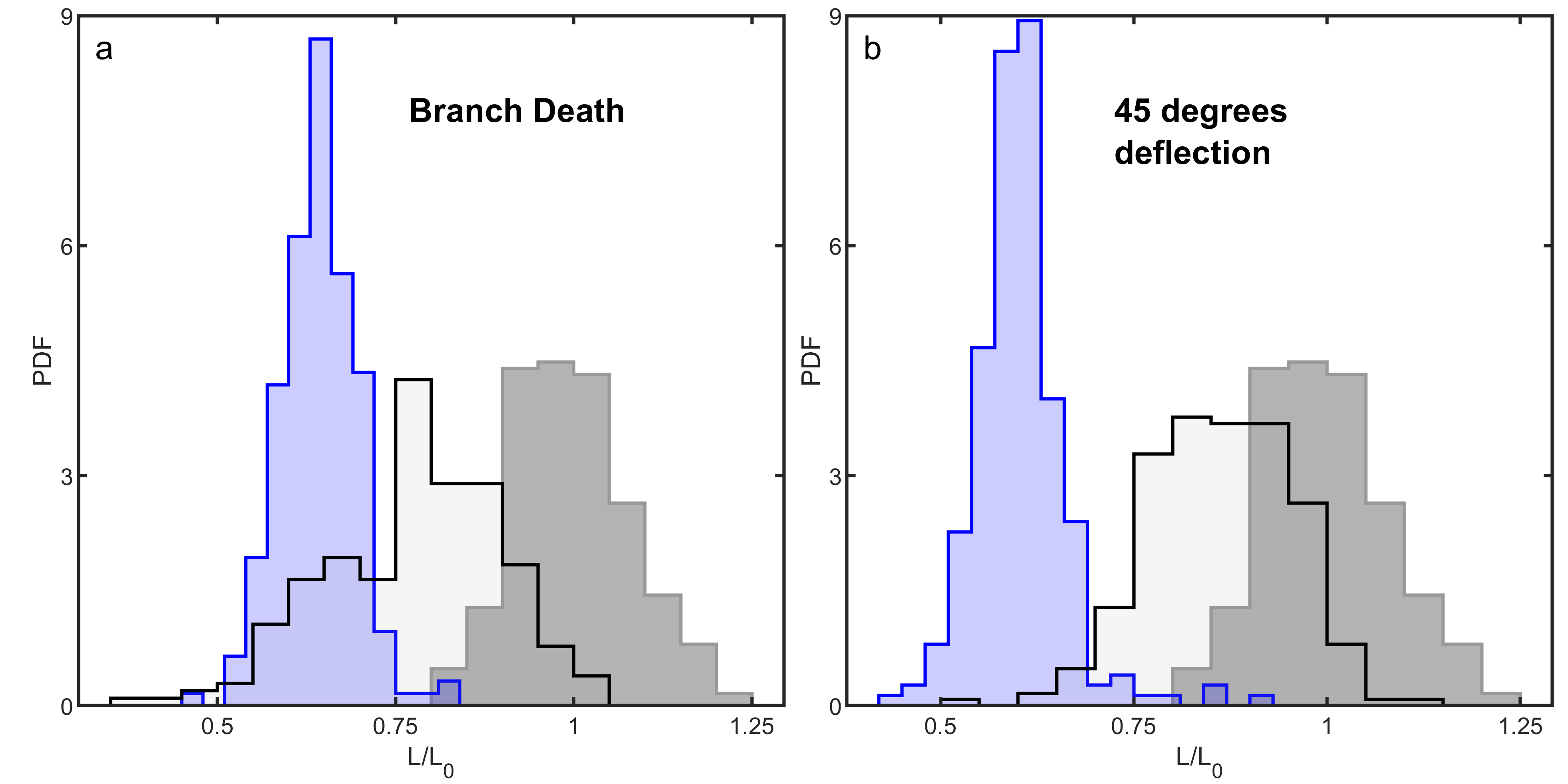}
    \caption{Growing and pushing simulations for situations where, instead of deflecting 90 degrees when encountering an existing branch, hyphal tips either died (left) or deflected by 45 degrees (right). In gray is the distribution of closest center of mass distance when trying to push trees as close as possible. Blue is the distribution of center of masses after growth. Black line is the result of taking the grown trees and trying to re-push them together.}
    \label{SF_03}
\end{figure*}

We also tested simulations where branches did not deflect but instead were terminated (deleted from the hyphal tip list), and others where branches deflected by $45$ degrees instead of 90 degrees. In both cases, trees that were grown near each other did not recover their original positions (Figure \ref{SF_03}).

\subsection*{Mechanical agitation of branching trees}
To mechanically agitate the trees, we could apply global rotations and translations to all points in the tree list. We then iteratively agitated via a combination of a translation of 2 units in a random direction and a rotation of 2 degrees around a random axis, checked for alpha shape collisions between two trees, and accepted or rejected the agitation. Each agitation was accepted if either (i) it resulted in a smaller overlapping volume than in the previous agitation step, or (ii) with a Boltzmann probability $p=e^{-\Delta V/T}$ where $\Delta V = V_i - V_{i-1}$ is the difference between current and previous overlapping volumes, with annealing temperature $T=10$ units, which was generally restrictive (moves that increased overlapping volume were rejected $64\%$ of the time). If the move was accepted, it was stored as the current state of the tree. If it was rejected, we restored the last known state of the tree and continued the agitation process.

\begin{figure*}
    \includegraphics[width=.8\linewidth]{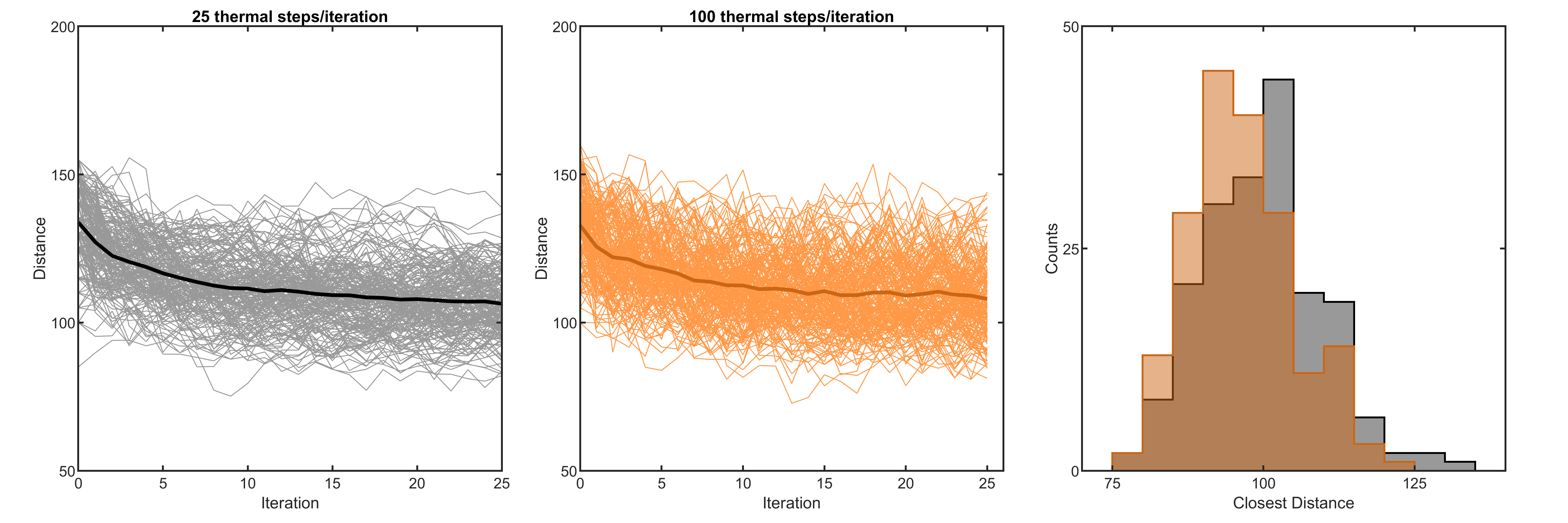}
    \caption{Testing the algorithm for pushing/agitating trees as close as possible. Left: center of mass distance vs. iteration number for 300 different simulated trajectories, where each iteration cycles (a) 25 random ``kicks'' combining a random translation and random rotation, then (b) an external force that pushes the trees together until they collide. Dark line is the mean of the different simulations. Middle: the same plot where there are 100 random kicks instead of 25 random kicks per iteration. Right: The distribution of closest center of mass distance across the 300 simulations for both cases.}
    \label{SF_04}
\end{figure*}

Agitation experiments generally had two stages. First, the trees were pushed together until they collided. Then, trees were agitated as described in the above paragraph for a set number of iterations. Then, the trees were again pushed together, and again agitated, etc. The number of thermal agitation steps was 25 used in the main text; we also tested using 100 thermal steps, which we found did not change the results (Supplement Figure \ref{SF_04}). Further, we found diminishing returns for continued cycles of pushing and jiggling; we decided to terminate the simulations after 25 cycles based on these results (Supplemental Figure \ref{SF_04}).

\subsection*{Controlling for morphological differences between grown and pushed trees}
We next controlled for morphological differences between trees that are grown separately and trees that are grown close to each other. We collected pairs of trees that were grown at a close separation distance, and separated them by a far distance ($2 L_0$). We then used the same pushing agitation procedure outlined above to push them together. Rather than achieving their original, close configuration ($\sim 50$ units) once again, we found that trees achieved an average minimum distance of $87.5\pm 8.1$ ($0.91L_0$) units. This value is similar to the average minimum distance achieved by trees undergoing solely agitation ($p=0.29$, $z=-1.06$), indicating that the close configuration is truly inaccessible to agitated trees. In none of the $134$ instances did trees reach the same or closer distance through agitation than they did through growth.

\subsection*{Measuring entanglement for a range of different growth geometries}
To explore the probability of entanglement for growing hyphal trees of different geometric properties, we grew trees nearby and then ``dragged'' them apart to measure collisions. First, we grew one tree in isolation with a selected value of $\Lambda$ and $\theta$ that define its geometric properties as described above, for a length of time $B=4$ which means that growth was truncated at $4$ branching events. The farthest reach of this tree was measured as $r_{max} = \textbf{max}(|r-\bar r|)$, where $r$ is a list of the Cartesian coordinates of the tree volume, $\bar r$ is the center of mass position of the tree, and $|...|$ denotes the vector magnitude. A random point on the surface of a sphere with radius $r_{max}$ was selected to be the initial seedpoint of a second tree, which was then grown with the same values of $\Lambda$ and $\theta$, and for a length of time $B$ which was varied from $3$ to $6$.

100 cases of each growth time $B$, and geometric property pair ($\Lambda,\theta$) were simulated. After growth, the simulations were checked for entanglement via a drag experiment. First, the centers of mass ($\bar r_1$ and $\bar r_2$) were found for both trees. The vector $r_{pull}=\bar r_2-\bar r_1$ pointing from tree 1 to tree 2 was defined as the dragging axis. Then, the second tree was pulled along that axis away from tree 1 in steps of length $2$ simulation units, until they were completely separated. At each step, the overlap of the two alpha shapes was measured. Entanglement was defined to occur when there was a peak in the overlap volume greater than a threshold value of $T=\pi*d^3$, or the volume of a cylinder with radius $d/2$ and height $d$, where $d$ is the diameter of the branches. This drag experiment was repeated for all 100 simulations of each geometric and growth time value, and the proportion of entangling simulations was measured.

\subsection*{Tunneling hyphae}
We also used a modified version of the branched-tree simulations to test predictions of our tunneling model. In these simulations, hyphae began as one hyphal tip growing in the positive x-direction. Then, hyphal tips would extend and periodically split, as described above. For the tunneling simulations, the hyphal tips interacted with a porous medium rather than with another tree. Additionally, the number of simultaneous hyphal tips was capped at 500 for computational speed. When the number of tips exceeded 500, some tips were pruned (deleted from the list) until the number of tips was below 500.

The porous medium used for these simulations was a sub-sampled block of data taken from the SEM experiments of the snowflake yeast. This sub-sampled, voxelized block of data contained yeast cells at a packing fraction of $\phi=0.38$. The block was truncated at various depths to obtain different porous medium widths. To obtain different volume fractions, the voxelized dataset was eroded or dilated (using MatLab algorithms for binary image erosion and dilation) with a cubic kernel. For different sized kernels, we eroded or dilated to different volume fractions, allowing for densities ranging from $0.004$ to $0.711$, calculated by dividing the number of voxels considered ``on'' by the total number of voxels within the sample block. These densities were normalized by the bond percolation critical value for cubic lattices in three dimensions, $p_c=0.7530$ \cite{Sykes1964, Gaunt1983}.

\bibliography{references}

\end{document}